\documentclass[twocolumn,showpacs,prc,superscriptaddress,nofootinbib]{revtex4}          
\newcommand{ \be }{\begin{equation}}       
\newcommand{ \ee }{\end{equation}}       
\newcommand{ \bea }{\begin{eqnarray}}       
\newcommand{ \eea }{\end{eqnarray}}

\newcommand{ \mean }[1]{\left\langle #1 \right\rangle}   
\newcommand{ \etal }{{\it et al.}}   
\usepackage{color}

\usepackage{graphicx} % Include figure files
\usepackage{fixmath} % bold math headlines          
\usepackage{dcolumn} % Align table columns on decimal point          
\usepackage{bm} % bold math          
\usepackage{multirow}
 
\begin{document}          
\title{       
%\begin{flushright}  
%{\small \sl version 2,  
%\today \\  
%} 
%\end{flushright}
Event-shape-engineering study of charge separation in heavy-ion collisions 
} 

\author{Fufang Wen}
\affiliation{Department of Physics and Astronomy, University of California, Los Angeles, California 90095, USA} 
\author{Jacob Bryon}
\affiliation{Department of Physics, University of California, Berkeley, California 94720, USA} 
\author{Liwen Wen}
\affiliation{Department of Physics and Astronomy, University of California, Los Angeles, California 90095, USA}
\author{Gang Wang}
\affiliation{Department of Physics and Astronomy, University of California, Los Angeles, California 90095, USA}

\begin{abstract}
Recent measurements of charge-dependent azimuthal correlations in high-energy heavy-ion collisions have indicated 
charge-separation signals perpendicular to the reaction plane,
and have been related to the chiral magnetic effect (CME).
However, the correlation signal is contaminated with the background caused by the collective motion (flow) of the collision system, and 
an effective approach is needed to remove the flow background from the correlation.
We present a method study with simplified Monte Carlo simulations and a multi-phase transport model, 
and develop a scheme to reveal the true CME signal via the event-shape engineering with the flow vector of the particles of interest. 
\end{abstract} 
 \pacs{25.75.Ld}
  
\maketitle  

%%%%%%%%%%%%%%%%%%%%%%%%%%%%%%%%%%%%%%%%%%%%%%%%%%%%%%%%%%%%%%
\section{Introduction}
\begin{figure}[t]
  \includegraphics[width=0.40\textwidth]{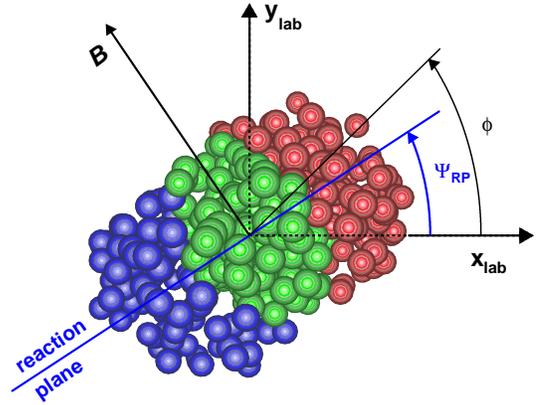}
  \caption{
    (Color online) Schematic depiction of the transverse plane for a collision of two heavy ions
        (the left one emerging from and the right one going into the page).
        Particles are produced in the overlap region (green-colored nucleons). The azimuthal angles
        of the reaction plane and a produced particle used in the three-point correlator, $\gamma$, are depicted here.
}
    \label{fig:Overlap}
\end{figure}

The hot, dense, and deconfined nuclear medium formed in high-energy heavy-ion collisions could bear a non-zero axial chemical potential ($\mu_5$), which characterizes the imbalance of right-handed and left-handed fermions in the system.
The chirality imbalance 
may be created locally in the medium through various mechanisms on an event-by-event basis 
(e.g. topological fluctuations in the gluonic sector, glasma flux tubes, and fluctuations in the quark 
sector)~\cite{Kharzeev_NPA2008,Kharzeev_PLB2006,Kharzeev_NPA2007,Kharzeev_PLB2002,Yin_PRL2015,Kharzeev_PRL2010}.
In a noncentral collision, a strong magnetic field ($B \sim 10^{15}$~T) can be produced (mostly by energetic spectator 
protons)~\cite{Kharzeev_PLB2006,Kharzeev_NPA2007}, and will induce an electric current along $\overrightarrow{B}$ in chiral domains,
$\overrightarrow{J_e} \propto \mu_5\overrightarrow{B}$, according to the chiral magnetic effect 
(CME)~\cite{Kharzeev_NPA2008,Kharzeev_PLB2006}. 
On average, $\overrightarrow{B}$ is perpendicular to the reaction plane (${\rm \Psi_{RP}}$) that contains the impact
parameter and the beam momenta, as depicted in Fig.~\ref{fig:Overlap}.
Hence the CME will manifest a charge transport across the reaction plane.

In the presence of the CME and other modes of collective motions in heavy-ion collisions, we can Fourier decompose the azimuthal distribution of produced particles of given transverse momentum ($p_T$) and rapidity:
\begin{eqnarray}
\frac{dN_{\alpha}}{d\phi} &\propto& 1 + 2v_{1,\alpha}\cos(\Delta\phi) + 2v_{2,\alpha}\cos(2\Delta\phi) + ... \nonumber \\
&+& 2a_{1,\alpha}\sin(\Delta\phi),
\label{equ:Fourier_expansion}
\end{eqnarray}
where $\phi$ is the azimuthal angle of a particle, and $\Delta\phi = \phi - {\rm \Psi_{RP}}$. 
Here the subscript $\alpha$ ($+$ or $-$) denotes the charge sign of the particle.
Conventionally $v_1$ is called ``directed flow" and $v_2$ ``elliptic flow"~\cite{ArtSergei}.
If caused by the CME, the parameter $a_1$ will be
nonzero with $a_{1,-} = -a_{1,+}$.
However, from event to event, the signs of the $\mu_5$ are equally likely, and 
the signs of finite $a_{1,+}$ and $a_{1,-}$ will flip accordingly,
leading to $\langle a_{1,+} \rangle = \langle a_{1,-} \rangle = 0$ on event average.
One therefore has to search for the CME with charge-separation \textit{fluctuations} 
perpendicular to the reaction plane, e.g., with a three-point correlator~\cite{Voloshin:2004vk},
$\gamma \equiv \langle \langle \cos(\phi_\alpha + \phi_\beta -2{\rm \Psi_{RP}}) \rangle_{\rm P}\rangle_{\rm E}$,
where the averaging is done over all particles in an event and over all events.
In practice, the reaction plane is approximated with the ``event plane" ($\rm \Psi_{EP}$) reconstructed with measured particles,
and then the measurement is corrected for the finite event plane resolution.

The expansion of the $\gamma$ correlator,
\begin{eqnarray}
&& \mean{\mean{\cos(\phi_{\alpha}+\phi_{\beta}-2{\rm \Psi_{RP}})}} \nonumber \\
&=& \mean{\mean{\cos(\Delta\phi_{\alpha})\cos(\Delta\phi_{\beta}) -
\sin(\Delta\phi_{\alpha})\sin(\Delta\phi_{\beta})}} \nonumber \\
&=& [\mean{v_{1,\alpha}v_{1,\beta}} + B_{\rm IN}] -[\mean{a_{1,\alpha}a_{1,\beta}} + B_{\rm OUT}], \label{eq:ThreePoint}
\end{eqnarray}
reveals the difference between the {\it in-plane} and {\it out-of-plane} projections of the correlations. 
The first term ($\mean{v_{1,\alpha}v_{1,\beta}}$) in the expansion provides a baseline unrelated to the magnetic field.
The background contribution ($B_{\rm IN}-B_{\rm OUT}$) is suppressed to a level close to 
the magnitude of $v_2$~\cite{Voloshin:2004vk}.
Previous measurements from STAR and ALICE Collaborations reported a robust charge-separation signal from 
the opposite- and same-charge $\gamma$ correlators ($\gamma_{\rm OS}>\gamma_{\rm SS}$) in Cu+Cu/Au+Au/Pb+Pb/U+U collisions 
with the center-of-mass energy from 19.6 GeV to 2.76 TeV~\cite{LPV_STAR1,LPV_STAR2,LPV_STAR3,LPV_STAR4,LPV_ALICE,LPV_UU};
see Ref.~\cite{Jinfeng} for a recent review of the experimental results.
However, the apparent charge separation is still contaminated with the $v_2$-related background.
For example, owing to elliptic flow, there are more clusters flying in-plane than out-of-plane,
which is irrelevant to the CME, and the decays of the clusters (into particles with opposite charges)
will contribute to the charge separation across the reaction plane.
Similar scenarios have been taken into consideration~\cite{Pratt2010,Flow_CME,PrattSorren:2011} where elliptic flow is coupled with transverse momentum conservation (TMC) or 
local charge conservation (LCC).

Flow backgrounds could be potentially removed via the event-shape-engineering (ESE) approach~\cite{Voloshin:2010ut,Schukraft:2012ah},
whereby {\it spherical} events or sub-events are selected,
so that the particles of interest therein carry zero $v_2$.
A previous attempt was made with a charge-separation observable roughly equivalent to $\gamma$, as
a function of event-by-event ``observed $v_2$"~\cite{LPV_STAR5}.
However, several issues in this implementation prevent a clear interpretation of the result.
In Sec~\ref{sec:ese}, we offer a few caveats in the practice of the ESE,
and develop an effective scheme to restore the CME signal with simplified Monte Carlo simulations and
a hybrid transport model (AMPT)~\cite{ampt1,ampt2}.

%=================================================================
\section{Event-shape engineering}
\label{sec:ese}
A valid ESE approach requires three key components.
First, a direct handle on the event shape is needed to truly reflect
the ellipticity of the particles of interest in each event.
This is not a trivial requirement,
as will be demonstrated later with a simple Monte Carlo simulation.
Second, the flow background has to vanish when the event-shape handle is turned to the zero-flow mode.
Here the AMPT model, which contains only backgrounds but no signals, will serve the illustration purpose.  
Third, the event selection by turning a proper handle should not introduce an artificial background, or if it does, the impact has to be under control. 
To study this effect, we will again use the simple simulation that contains only signals but no backgrounds.

\subsection{Handle on event shape}
\label{sec:handle}
In the analyses involving the reaction plane, it is common practice to divide particles in each event into 
two or more sub-events, and each sub-event has its own flow vector, $\overrightarrow{q}=(q_x,q_y)$:
\bea
q_x &\equiv& \frac{1}{\sqrt{N}} \sum_i^N \cos(2\phi_i) \label{qx}  \\
q_y &\equiv& \frac{1}{\sqrt{N}} \sum_i^N \sin(2\phi_i). \label{qy}
\eea
For example, a flow analysis involves two
sub-events, A and B, and reconstructs the event plane ${\rm \Psi_{EP}^B}$ (the azimuthal angle of 
$\overrightarrow{q}^{\rm B}$),
and then correlates particles in A with ${\rm \Psi_{EP}^B}$~\cite{ArtSergei}:
\be
v_{2}^{\rm observe} \equiv \langle\langle \cos[2(\phi^{\rm A} - {\rm \Psi_{EP}^B})] \rangle_{\rm P}\rangle_{\rm E}.
\ee
The true $v_2$ of particles in A (with respect to the true reaction plane) 
\be
v_2^A \equiv \langle\langle \cos[2(\phi^{\rm A} - {\rm \Psi_{RP}})] \rangle_{\rm P}\rangle_{\rm E}
\label{factorization}
\ee
will be obtained with $v_2^{\rm observe} / R^{\rm B}$, where $R^{\rm B}$ is 
the event plane resolution of ${\rm \Psi_{EP}^B}$~\cite{ArtSergei}
\be
R^{\rm B}\equiv \langle \cos[2({\rm \Psi_{EP}^B}-{\rm \Psi_{RP}})] \rangle_{\rm E}.
\label{resolution}
\ee
The single-bracket means the averaging over events.
The sub-event-plane method described above has been extensively used and proven by many to be valid on event-ensemble average.

In the ESE implementation, it is tempting to introduce an event-by-event ``$v_2$" observable,
\be
v_{2,{\rm ebye}}^{\rm observe} \equiv \langle \cos[2(\phi^{\rm A} - {\rm \Psi_{EP}^B})] \rangle_{\rm P},
\label{v2ebye}
\ee
and to define ``spherical" class-A sub-events with the condition that $v_{2,{\rm ebye}}^{\rm observe}=0$,
as was done in Ref.~\cite{LPV_STAR5}. However, zero $v_{2,{\rm ebye}}^{\rm observe}$
does not necessarily mean that particles in A have zero $v_{2}^{\rm A}$.
%Here $v_{2,{\rm ebye}}^{\rm A} \equiv \langle \cos[2(\phi^{\rm A} - {\rm \Psi_{RP}})] \rangle_{\rm P}$ 
%is elliptic flow of sub-event A with respect to the true reaction plane.

\begin{figure}[!htb]
  \centering
  \includegraphics[width=.5\textwidth]{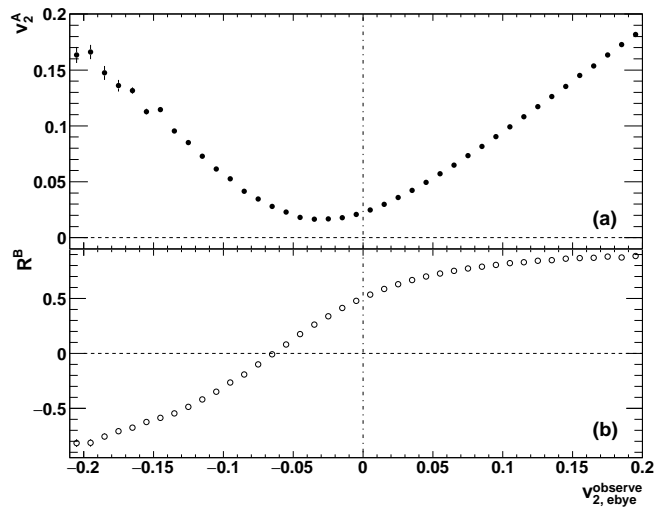}
  \caption{The true elliptic flow $v_{2}^{\rm A}$ (as in Eq.~\ref{factorization}) (a) and the true event plane resolution 
	$R^{\rm B}$ (as in Eq.~\ref{resolution}) (b) as functions of $v_{2,{\rm ebye}}^{\rm observe}$ (as in Eq.~\ref{v2ebye}), from simplified Monte Carlo simulations.
}
  \label{fig:v2_v2obs}
\end{figure}

We study the relationship between $v_{2}^{\rm A}$ and $v_{2,{\rm ebye}}^{\rm observe}$ with a simplified Monte Carlo simulation. In each event, the azimuthal angle of each particle 
has been assigned randomly according to the distribution of Eq.~(\ref{equ:Fourier_expansion}).
In this Monte Carlo simulation, the only nonzero harmonics are $v_2 = 5\%$ and $|a_{1,\pm}|=2\%$, and
non-flow effects such as TMC, LCC and resonance decay have not been implemented.
In other words, there are only elliptic flow and the charge separation due to the CME, but no background contributions.
Each of the 10 million simulated events contains 400 charged particles, with 200 positively charged and 200 negatively charged.

Figure~\ref{fig:v2_v2obs} presents the simulation results of $v_{2}^{\rm A}$
(a) and $R^{\rm B}$ (the event plane resolution of ${\rm \Psi_{EP}^B}$) (b) as functions of $v_{2,{\rm ebye}}^{\rm observe}$.
The $v_{2}^{\rm A}$ and $R^{\rm B}$ values,
respectively, have been averaged over events within the same $v_{2,{\rm ebye}}^{\rm observe}$ bin.
The upper panel displays an interesting U-shape in $v_{2}^{\rm A}$
vs $v_{2,{\rm ebye}}^{\rm observe}$,
with the minimum above zero. This means that truly spherical class-A sub-events can never be selected no matter how 
the $v_{2,{\rm ebye}}^{\rm observe}$ handle is turned. Or at least, the sphericity of particles in A depends on the choice of the 
{\it beholder} (${\rm \Psi_{RP}}$ or ${\rm \Psi_{EP}^B}$), if the event-by-event ``$v_2$" observable is the selection criterion.
Worse still is the fact exhibited in the lower panel that $R^{\rm B}$
strongly depends on $v_{2,{\rm ebye}}^{\rm observe}$, and could become negative.
This makes it highly nontrivial to correct for the event plane resolution, when the same ${\rm \Psi_{EP}^B}$ is
used to calculate $v_{2}^{\rm A}$ or $\gamma^{\rm A}$ differentially as a function of $v_{2,{\rm ebye}}^{\rm observe}$.
In reality, the negative $R^{\rm B}$ values are hardly extractable. 
The more disturbing caveat comes from the combined information from both panels:
\be
v_{2}^{\rm A} \neq v_{2,{\rm ebye}}^{\rm observe} / R^{\rm B},
\label{no_factorization}
\ee
for any $v_{2,{\rm ebye}}^{\rm observe}$ bin.
Therefore, even with the knowledge of $R^{\rm B}$, it is unlikely to restore the value of an
observable with respect to the true reaction plane in the $v_{2,{\rm ebye}}^{\rm observe}$-binning scheme. 
The factorization assumption underlying the correction for the event plane resolution is valid on
event-ensemble average, but breaks down on the $v_{2,{\rm ebye}}^{\rm observe}$ basis.

\begin{figure}[!htb]
  \centering
  \includegraphics[width=.5\textwidth]{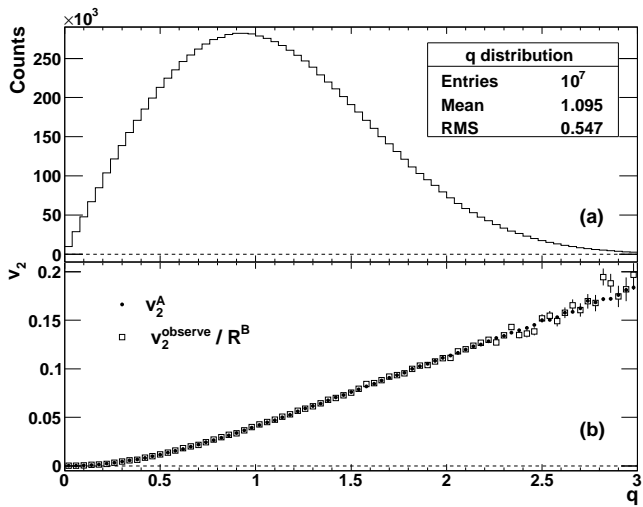}
  \caption{The distribution of $q$ (a), and the true elliptic flow $v_{2}^{\rm A}$ and 
	the corrected $v_{2}^{\rm observe}$ as functions of $q$ (b), from Monte Carlo simulations.
}
  \label{fig:v2_Q}
\end{figure}

A good handle on event shape should directly reflect the sphericity property of the particles of interest, independent of the beholder.
One candidate is $q$, the magnitude of the flow vector (as in Eqs.~(\ref{qx}) and (\ref{qy})) reconstructed with particles in the sub-event A.
By definition, $q$ has no explicit contributions from the sub-event B or the reaction plane.
In reality, there could be implicit correlations between $q$ and ${\rm \Psi_{EP}^B}$ due to flow fluctuations,
which will be discussed in Sec~\ref{sec:ampt}. 
Figure~\ref{fig:v2_Q}(b) presents the Monte Carlo simulation results of the true elliptic flow $v_{2}^{\rm A}$ 
and the corrected observable $v_{2}^{\rm observe}$ as functions of $q$.
On the $q$ basis, the correction for the event plane resolution is valid, and both $v_{2}$ values
approach zero at vanishing $q$. Note that the success in a simple simulation does not guarantee the success in 
real-data analyses, which can be complicated by various realistic factors, but if an approach 
(like the $v_{2,{\rm ebye}}^{\rm observe}$ basis) fails even in a simple simulation, it should definitely be avoided
in data analyses.

\begin{figure}[!htb]
  \centering
  \includegraphics[width=.5\textwidth]{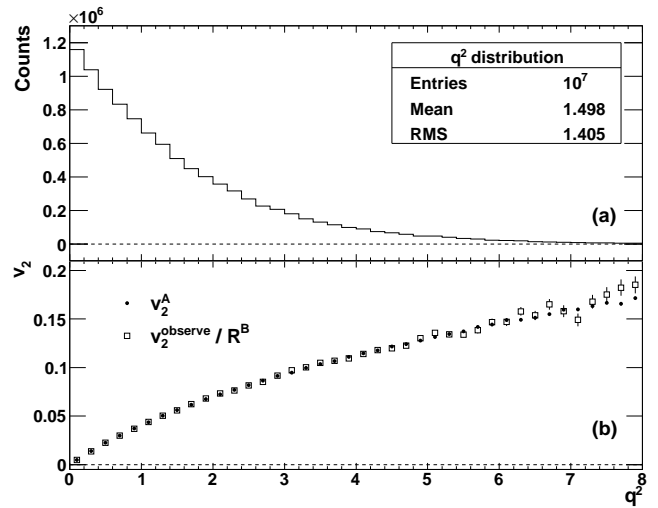}
  \caption{The distribution of $q^2$ (a), and the true elliptic flow $v_{2}^{\rm A}$ and
        the corrected $v_{2}^{\rm observe}$ as functions of $q^2$ (b), from Monte Carlo simulations.
}
  \label{fig:v2_Q2}
\end{figure}
$q$ is a good handle on event shape, but not as good as $q^2$.
$q^2=0$ implies $q=0$, so $q^2$ naturally inherits the capability of selecting spherical sub-events in terms of the second harmonic.
Moreover, Fig.~\ref{fig:v2_Q2}(b) displays a close-to-linear relationship between $v_{2}^{\rm A}$ 
($v_{2}^{\rm observe}/R^{\rm B}$) and $q^2$ at low $q^2$, which makes it more reliable to project $\gamma_{\rm A}$ to $q^2=0$, to remove $v_2$-related backgrounds.
Another advantage of $q^2$ over $q$ lies in their distributions, shown in Fig.~\ref{fig:v2_Q}(a) and Fig.~\ref{fig:v2_Q2}(a).
The $q$ distribution peaks around unity, and rapidly drops on both sides.
This feature means lower statistics towards lower $q$, so the projection of an event-by-event observable 
to $q=0$ becomes unstable. On the other hand, the $q^2$ distribution is shifted in phase space towards zero,
facilitating a statistically robust projection to zero $q^2$. 

\subsection{Disappearance of background}
\label{sec:ampt}
The AMPT model~\cite{ampt1,ampt2} is a realistic event generator that has been widely used to describe experimental data.
The string melting version of AMPT~\cite{ampt2,ampt3} reasonably well reproduces particle spectra and elliptic flow
in Au+Au collisions at $\sqrt{s_{\rm NN}}=200$ GeV and Pb+Pb collisions at 2.76 TeV~\cite{ampt4}.
The CME is not included in AMPT, which simplifies the background study.
10 million AMPT events are generated for 200 GeV Au+Au collisions.
Each event is divided into three sub-events according to particle pseudorapidity ($\eta$):
sub-event A contains particles of interest with $|\eta|<1.5$,
and sub-event B1 (B2) provides a sub-event plane using particles with $1.5<\eta<4$ ($-4<\eta<-1.5$). 
${\rm \Psi_{EP}^{B1}}$ and ${\rm \Psi_{EP}^{B2}}$ are separately used to calculate
$v_2$ or $\gamma$, and the two sets of results are combined to achieve better statistics.
The corresponding sub-event plane resolution is obtained via correlation:
\be
R^{\rm B1(B2)}\equiv\sqrt{\langle\cos[2({\rm \Psi_{EP}^{B1}}-{\rm \Psi_{EP}^{B2}})]  \rangle_{\rm E}}.
\label{res_B1B2}
\ee

\begin{figure}[!htb]
  \centering
  \includegraphics[width=.5\textwidth]{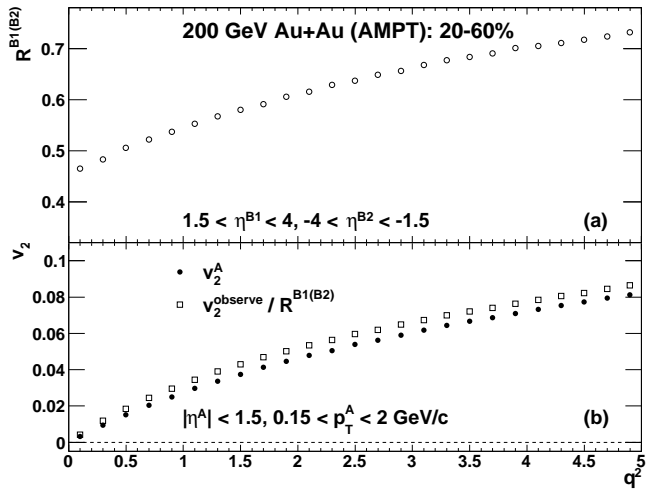}
  \caption{The sub-event plane resolution (as in Eq.~\ref{res_B1B2}) (a), and the true elliptic flow $v_{2}^{\rm A}$ and the corrected $v_{2}^{\rm observe}$ (b), as functions of $q^2$, from AMPT simulations.
}
  \label{fig:AMPT_v2_Q2}
\end{figure}

Figure~\ref{fig:AMPT_v2_Q2} shows the sub-event plane resolution (a), and the true elliptic flow $v_{2}^{\rm A}$ 
and the corrected $v_{2}^{\rm observe}$ (b), as functions of $q^2$, from AMPT simulations of $20-60\%$
Au+Au collisions at $\sqrt{s_{\rm NN}} = 200$ GeV.
Unlike the simplified Monte Carlo simulation in which $R^{\rm B}$ is constant over $q^2$, 
AMPT events involve flow fluctuation that causes a positive correlation in flow
between sub-events in a same event, and as a result, $R^{\rm B1(B2)}$ increases with $q^2$.
The lower panel displays a discrepancy between $v_{2}^{\rm A}$ and the corrected $v_{2}^{\rm observe}$,
which is not a sign of the breakdown of the underlying factorization assumption,
but due to the difference between the reaction plane and the participant plane~\cite{participant_plane},
in terms of non-flow and flow fluctuation.
It matters more that both $v_2$ values decrease with decreasing $q^2$, and drop to $(0,0)$.

\begin{figure}[!htb]
  \centering
  \includegraphics[width=.5\textwidth]{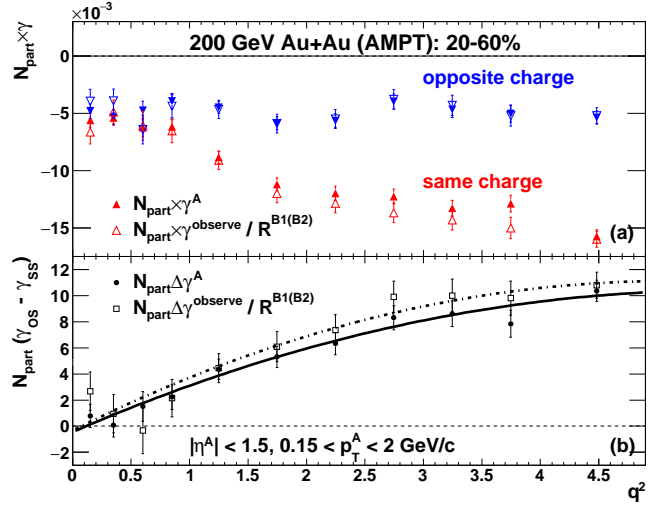}
  \caption{(Color online) $N_{\rm part}\times\gamma$ (a) and $N_{\rm part}\Delta\gamma$ (b) 
as functions of $q^2$, from AMPT simulations.  The full (open) symbols represent results obtained
with the true reaction plane (reconstructed event plane, with correction for the event plane resolution).
The solid (dashed) line in the lower panel is a $2^{\rm nd}$-order polynomial fit of the full (open) data points.
}
  \label{fig:AMPT_gamma_Q2}
\end{figure}

Figure~\ref{fig:AMPT_gamma_Q2}(a) presents the $\gamma$ correlators multiplied by the number 
of participating nucleons, $N_{\rm part}$, as functions of $q^2$, for $20-60\%$ AMPT events of Au+Au collisions at 200 GeV.
Here $N_{\rm part}$ is used to compensate for the dilution effect due to the later-stage rescattering~\cite{Ma:2011uma}.  
For both the same-charge and the opposite-charge correlators, the true $\gamma^{\rm A}$ and the corrected
$\gamma^{\rm observe}$ are consistent with each other within the statistical uncertainties.
This indicates that compared with $v_2$, $\gamma$ better supports the validity of the correction for the event plane resolution, and is less sensitive to non-flow or flow fluctuation.
At larger $q^2$, the opposite-charge correlators are above the same-charge correlators,
suggesting a finite flow-related background. The opposite- and same-charge correlators converge at small $q^2$.
The lower panel shows $N_{\rm part}\Delta\gamma \equiv N_{\rm part}(\gamma_{\rm OS}-\gamma_{\rm SS})$ vs $q^2$,
and again, the two observables seem to coincide. 
$2^{\rm nd}$-order polynomial fits to both observables yield small intercepts that are consistent with zero:
$(-4.5\pm6.7)\times10^{-4}$ for $N_{\rm part}\Delta\gamma^{\rm A}$ 
and $(-3.3\pm10.6)\times10^{-4}$ for $N_{\rm part}\Delta\gamma^{\rm observe}/R^{\rm B1(B2)}$.
The finite $\Delta\gamma$ values in AMPT events are solely due to background contributions, 
so the disappearance of background is demonstrated when the ``correctable" observable ($\Delta\gamma$) is projected to zero $q^2$. 
Here the $2^{\rm nd}$-order polynomial fits only serve for illustration purposes, and the optimal projection scheme 
is subject to the details of the measured $\Delta{\gamma}(q^2)$.

\subsection{Artificial signal/background}
\label{sec:artificial}
The study of the physical relationship between two observables has to obviate
the mathematical correlation between their definitions. For example, when $\gamma_{\rm ebye}$ is plotted against 
$v_{2,{\rm ebye}}$ (for simplicity, both are obtained with the true reaction plane), 
a finite slope often exists even if there are no explicit
physical correlations between the two. This can be understood by expanding $\gamma_{\rm ebye}$:
\bea
\gamma_{\rm ebye} &\equiv& \mean{\cos(\phi_{\alpha}+\phi_{\beta}-2{\rm \Psi_{RP}})}_{\rm P} \nonumber \\
&=& \mean{\cos[(\phi_{\alpha}-\phi_{\beta})+2(\phi_{\beta}-{\rm \Psi_{RP}})]}_{\rm P} \nonumber \\
&=& \mean{\cos(\phi_{\alpha}-\phi_{\beta})\cos[2(\phi_{\beta}-{\rm \Psi_{RP}})]}_{\rm P}  \nonumber \\
&& - \mean{\sin(\phi_{\alpha}-\phi_{\beta})\sin[2(\phi_{\beta}-{\rm \Psi_{RP}})]}_{\rm P} \nonumber \\
&\approx& 2\delta_{\rm ebye}v_{2,{\rm ebye}} + C,
\label{eq:artificial}
\eea
where $C$ is an event-ensemble-averaged quantity, and  
$\delta_{\rm ebye}\equiv\mean{\cos(\phi_{\alpha}-\phi_{\beta})}_{\rm P}$, is the two-particle correlation,
which contains various contributions such as $a_{1,\alpha}a_{1,\beta}$, resonance decay, TMC,
LCC, etc. $\delta$ is usually finite, leading to a finite apparent slope
in $\gamma_{\rm ebye}$ vs $v_{2,{\rm ebye}}$, which could be misinterpreted as a physical relationship.
The coefficient ``$2$" in front of $\delta_{\rm ebye}v_{2,{\rm ebye}}$
reflects the contributions from both the $\mean{\cos(...)\cos(...)}$ term
and the $-\mean{\sin(...)\sin(...)}$ term.
Without loss of generality, we apply the initial condition that on event-ensemble average
$\gamma_{\alpha\beta}$ is $-a_{1,\alpha}a_{1,\beta}$,
and thus the quantity $C$ becomes $-a_{1,\alpha}a_{1,\beta} - 2\delta v_2$.

\begin{figure}[!htb]
  \centering
  \includegraphics[width=.5\textwidth]{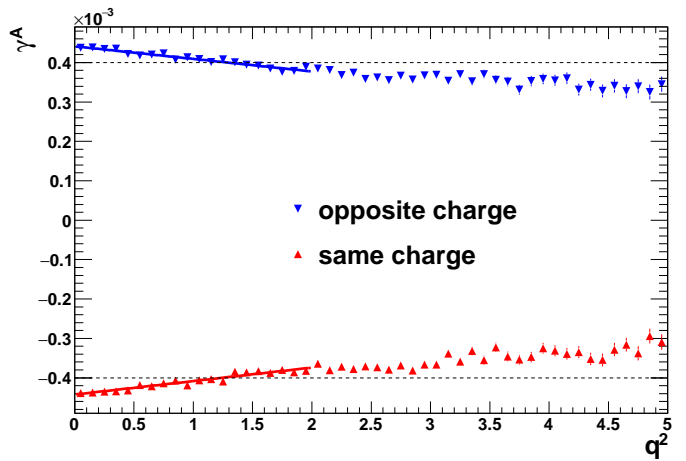}
  \caption{(Color online) $\gamma$ obtained with the true reaction plane as a function of $q^2$, from the Monte Carlo simulation.  
The solid lines are linear fits of the points.
}
  \label{fig:Toy_gamma_Q2}
\end{figure}

As discussed in Sec.~\ref{sec:ampt}, the flow background disappears when $q^2=0$, therefore any finite $\Delta\gamma$ 
signal at zero $q^2$ in experimental measurements will evidence a charge separation truly due to the CME.   
However, a finite signal at zero $q^2$ is not necessarily equal to the event-ensemble-averaged signal.
Fig.~\ref{fig:Toy_gamma_Q2} illustrates the artificial effect with the simplified Monte Carlo simulation,
where the input $|a_1|$ and $v_2$ are fixed at $2\%$ and $5\%$, respectively, and there are no backgrounds or explicit physical correlations between $|a_1|$ and $v_2$. 
The absence of background validates $\delta_{\rm SS} = - \gamma_{\rm SS}$ and $\delta_{\rm OS} = - \gamma_{\rm OS}$.
Although the event-ensemble average of $\gamma_{\rm SS}$ ($\gamma_{\rm OS}$)
is $-4\times10^{-4}$ ($4\times10^{-4}$), the apparent value at zero $q^2$ exaggerates the charge separation
by relative $2v_2$ ($10\%$ in this case), as described by Eq.~\ref{eq:artificial}.

\begin{figure}[!htb]
  \centering
  \includegraphics[width=.5\textwidth]{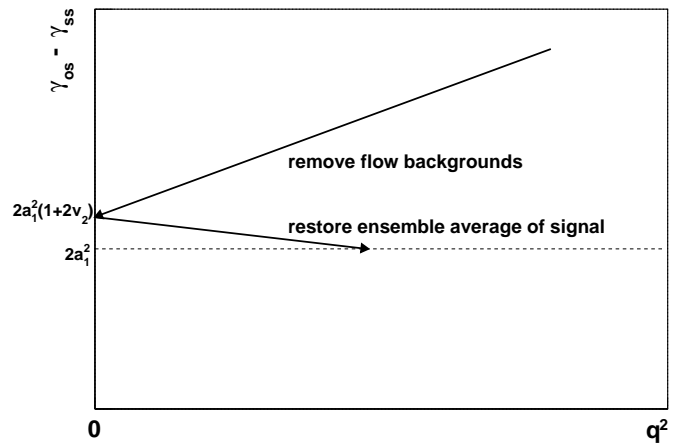}
  \caption{A schematic diagram of how to reveal the ensemble-averaged CME signal via the ESE.
}
  \label{fig:sketch}
\end{figure}

Figure~\ref{fig:sketch} sketches our proposal to experimentally reveal the true CME signal via the ESE.
First, the proper handle on event shape, $q^2$, is employed for the particles of interest in each event.
Second, the flow background is removed by projecting the charge-separation observable to zero $q^2$.
Third, the event-ensemble-averaged CME signal is restored from $\Delta\gamma_{\rm ebye}|_{q^2=0}/(1+2v_2)$.
This scheme is not unique to the $\gamma$ correlator, and could be applied to other similar observables, such as 
the modulated sign correlator (MSC)~\cite{LPV_STAR3} and the charge multiplicity asymmetry correlator (CMAC)~\cite{LPV_STAR5}.

%===============================
%==========================================================================

\section{Summary and discussion}
The experimental searches for the CME in heavy-ion collisions have aroused extensive attention,
and special efforts are warranted to disentangle the CME-induced charge-separation signal
from the flow-related backgrounds.
We have disclosed a few shortcomings of a previous attempt of the ESE with 
$v_{2,{\rm ebye}}^{\rm observe}$~\cite{LPV_STAR5}. 
The root cause lies in the fact that $v_{2,{\rm ebye}}^{\rm observe}$ is a correlation
between two symmetric sub-events, instead of a property of any sub-event.
Therefore, the selection of a given $v_{2,{\rm ebye}}^{\rm observe}$ value
triggers an event-shape bias in either sub-event, making neither suitable
to serve as an unbiased event plane.
For example, $v_{2,{\rm ebye}}^{\rm observe}=0$ implies that the two sub-event planes,
${\rm \Psi_{EP}^A}$ and ${\rm \Psi_{EP}^B}$, are $\pm45^{\circ}$ from each other, and neither sub-event has to be spherical.
In this case, when one sub-event (A) is beheld by the other (B), which is either $45^{\circ}$
or $-45^{\circ}$ away (not necessarily with equal possibility), the sense of being in-plane or out-of-plane
is impaired, because the two possible scenarios of ${\rm \Psi_{EP}^B}$ are perpendicular to each other.
As a result, the observed charge separation is artificially reduced at $v_{2,{\rm ebye}}^{\rm observe}=0$.

The magnitude of the flow vector, $q$, or even better, $q^2$,
emerges to be a good handle on event shape. 
$q$ or $q^2$ directly reflects the sphericity property of the sub-event of interest,
and zero $q$ selects spherical sub-events in the second harmonic.
$q^2$ is technically better than $q$, because $q^2$ is almost proportional to $v_2$ at low $q^2$,
and the $q^2$ distribution favors the projection of $\gamma$ to zero $q^2$.
The AMPT model has been exploited to verify the disappearance of flow backgrounds at zero $q^2$,
and simplified Monte Carlo simulations have been utilized to study the artificial correlations
in the ESE process.
Based on these findings, we have designed an effective recipe to experimentally remove flow backgrounds
and restore the event-ensemble average of the CME signal.

The ESE approach proposed in this work may be invalidated
by an extreme scenario~\cite{Jie}, where an in-plane-going resonance   decays into a positive particle at $45^{\circ}$ and a negative particle at $-45^{\circ}$.    
The ``charge separation" introduced this way will add to $\Delta\gamma$,
while the two daughters together have no contribution to $q$. In other words, even at $q=0$, the background from such flowing resonance will not completely vanish. However, the AMPT model includes realistic resonance yields, and does not display a significant effect at $q^2=0$, as shown in Fig.~\ref{fig:AMPT_gamma_Q2}. Therefore, we conclude that such an effect seems to be a rare case.

{\bf Acknowledgments:} We thank Huan Huang and other members of the UCLA Heavy Ion
Physics Group for discussions, and we are grateful to Ning Yu for the help with AMPT.
We also benefited from the fruitful discussions with Fuqiang Wang.
This work is supported by a grant (No. DE-FG02-88ER40424) from U.S.
Department of Energy, Office of Nuclear Physics.
%==========================================================================  
  

\begin{thebibliography}{99}          

%chirality
\bibitem{Kharzeev_NPA2008}
D. E. Kharzeev, L. D. McLerran and H. J. Warringa, Nucl. Phys. A {\bf 803}, 227 (2008)
\bibitem{Kharzeev_PLB2006}
D. Kharzeev, Phys. Lett. B 633, {\bf 260} (2006).
\bibitem{Kharzeev_NPA2007}
D. Kharzeev and A. Zhitnitsky, Nucl. Phys. A {\bf 797}, 67 (2007).
\bibitem{Kharzeev_PLB2002}
D. Kharzeev, A. Krasnitz and R. Venugopalan, Phys. Lett. B {\bf 545}, 298 (2002).
\bibitem{Yin_PRL2015}
I. Iatrakis, S. Lin and Y. Yin, Phys. Rev. Lett. {\bf 114}, 252301 (2015).
\bibitem{Kharzeev_PRL2010}
K. Fukushima, D. E. Kharzeev and H. J. Warringa, Phys. Rev. Lett. {\bf 104}, 212001 (2010).

% Art and Sergei's Flow method paper
\bibitem{ArtSergei} A.~M.~Poskanzer and S.~Voloshin, Phys.\ Rev.\ C {\bf 58}, 1671 (1998).

\bibitem{Voloshin:2004vk}
S. A. Voloshin, Phys. Rev. C {\bf 70}, 057901 (2004).

% STAR PRL parity paper
\bibitem{LPV_STAR1}
B. I. Abelev {\it et al.} [STAR Collaboration], Phys. Rev. Lett. {\bf 103}, 251601 (2009).
\bibitem{LPV_STAR2}
B. I. Abelev {\it et al.} [STAR Collaboration], Phys. Rev. C {\bf 81}, 54908 (2010).
\bibitem{LPV_STAR3}
L. Adamczyk {\it et al.} [STAR Collaboration], Phys. Rev. C {\bf 88}, 064911 (2013).
\bibitem{LPV_STAR4}
L. Adamczyk {\it et al.} [STAR Collaboration], Phys. Rev. Lett. {\bf 113}, 052302 (2014).

% ALICE PRL parity paper
\bibitem{LPV_ALICE}
B. I. Abelev \etal [ALICE Collaboration], \prl {\bf 110}, 012301 (2013).

%Gang's QM proceedings on U+U
\bibitem{LPV_UU}
Gang Wang {\it et al.} [STAR Collaboration], Nucl. Phys A {\bf 904-905}, 248c (2013).

%Review paper
\bibitem{Jinfeng}
D.E. Kharzeev, J. Liao, S.A. Voloshin and G. Wang, Prog. Part. Nucl. Phys. {\bf 88}, 1 (2016).

% Pratt momentum conservation
\bibitem{Pratt2010} S.~Pratt, S.~Schlichting and S.~Gavin, Phys.\ Rev.\ C {\bf 84}, 024909 (2011).
%kappa
\bibitem{Flow_CME}
A. Bzdak, V. Koch and J. Liao, Lect. Notes Phys. {\bf 871}, 503 (2013).
% Scott Pratt and Soren's Local charge conservation + v2 paper
\bibitem{PrattSorren:2011} S.~Schlichting and S.~Pratt, Phys.\ Rev.\ C {\bf 83}, 014913 (2011).
%\cite{Voloshin:2010ut} 
\bibitem{Voloshin:2010ut}  S.~A.~Voloshin, %``Testing the Chiral Magnetic Effect with Central U+U collisions,''
  Phys.\ Rev.\ Lett.\ {\bf 105}, 172301 (2010).
%\cite{Schukraft:2012ah} 
\bibitem{Schukraft:2012ah}  J.~Schukraft, A.~Timmins and S.~A.~Voloshin, %``Ultra-relativistic nuclear collisions: event shape engineering,''
  Phys.\ Lett.\ B {\bf 719}, 394 (2013).

\bibitem{LPV_STAR5}
L. Adamczyk {\it et al.} [STAR Collaboration], Phys. Rev. C {\bf 89}, 044908 (2014).

\bibitem{ampt1}
B. Zhang, C.M. Ko, B.-A. Li and Z.-W. Lin, Phys. Rev. C {\bf 61}, 067901 (2000).
\bibitem{ampt2}
Z.-W. Lin, C.M. Ko, B.-A. Li and B. Zhang, Phys. Rev. C {\bf 72}, 064901 (2005).
\bibitem{ampt3}
Z.-W. Lin and C.M. Ko, Phys. Rev. C {\bf 65}, 034904 (2002).
\bibitem{ampt4}
Z.-W. Lin, Phys. Rev. C {\bf 90}, 014904 (2014).

%Participant plane
\bibitem{participant_plane} J.~-Y.~Ollitrault, A.~M.~Poskanzer and S.~A.~Voloshin,
Phys.\ Rev.\ C {\bf 80}, 014904 (2009).

\bibitem{Ma:2011uma}
  G.~L.~Ma and B.~Zhang,
  %``Effects of final state interactions on charge separation in relativistic heavy ion collisions,''
  Phys.\ Lett.\ B {\bf 700}, 39 (2011).
\bibitem{PHOBOS1}
B. Alver {\it et al.} [PHOBOS Collaboration], Phys. Rev. C {\bf 83}, 024913 (2011).
\bibitem{PHOBOS2}
B.B Back {\it et al.} [PHOBOS Collaboration], Phys. Rev. C {\bf 72}, 051901(R) (2005).

\bibitem{Jie}
F. Wang and J. Zhao, Phys. Rev. C {\bf 95}, 051901(R) (2017).

\end{thebibliography}
\end{document}